\documentclass[11pt,draftcls,onecolumn]{IEEEtran}
\usepackage{amsfonts}
\usepackage{amssymb}
\usepackage{mathrsfs}
\usepackage{amsmath}
\usepackage{cite}
\usepackage{mathrsfs}
\usepackage[ruled,vlined]{algorithm2e}
\usepackage{tikz}
\usetikzlibrary{arrows,automata}
\usepackage[latin1]{inputenc}
\usepackage{verbatim}
\makeatletter

\newcommand{\Rmnum}[1]{\expandafter\@slowromancap\romannumeral #1@}
\makeatother

\newtheorem{thm}{Theorem}
\newtheorem{lem}[thm]{Lemma}
\newtheorem{eg}{Example}

\newtheorem{defn}{Definition}
\newtheorem{rem}[thm]{Remark}

\newcommand{\Fq}{\mathbb{F}_q}

\newcommand{\Rank}{{\mathrm{Rank}}}

\begin{document}

\title{Repairable Threshold Secret Sharing Schemes
\thanks{This research is supported by the National Key Basic Research Problem of China (973 Program Grant No. 2013CB834204), the National Natural Science Foundation of China (Nos. 61301137, 61171082) and the Fundamental Research Funds for Central Universities of China (No. 65121007).}}

\author{Xuan~Guang~\IEEEmembership{Member,~IEEE},
        Jiyong~Lu~\IEEEmembership{Student Member,~IEEE},
        and~Fang-Wei Fu
\thanks{X. Guang is with the School of Mathematical Sciences and LPMC, J. Lu and F.-W. Fu are with the Chern Institute of
Mathematics, Nankai University, Tianjin 300071, China (e-mail:
xguang@nankai.edu.cn, lujiyong@mail.nankai.edu.cn, fwfu@nankai.edu.cn).}}

\markboth{Repairable Threshold Secret Sharing Schemes}%
{}
{}
%


\maketitle

\begin{abstract}
In this paper, we propose a class of threshold secret sharing schemes with repairing function between shares without the help of the dealer, that we called repairable threshold secret sharing schemes. Specifically, if a share fails, such as broken or lost, it will be repaired just by some other shares. A construction of such repairable threshold secret sharing schemes is designed by applying linearized polynomials and regenerating codes in distributed storage systems. In addition, a new repairing rate is introduced to characterize the performance and efficiency of the repairing function. Then an achievable upper bound on the repairing rate is derived, which implies the optimality of the repair and describes the security between different shares. Under this optimality of the repair, we further discuss traditional information rate and also indicate its optimality, that can describe the efficiency of secret sharing schemes in the aspect of storage. Finally, by applying the minimum bandwidth regenerating (MBR) codes, our construction designs repairable threshold secret sharing schemes achieving both optimal repairing and information rates simultaneously.
\end{abstract}


%
\IEEEpeerreviewmaketitle

\section{Introduction}\label{intro}
\IEEEPARstart{I}{n} order to keep a secret (usually a cryptographic key) highly reliable and secure, Blakley \cite{Blakley-1979} and Shamir \cite{Shamir-1979} presented the concept of threshold secret sharing scheme in 1979 independently. In Shamir's $(t,n)$ threshold secret sharing scheme, a dealer shares a secret $S$ to $n$ participants $P_1,P_2,\cdots,P_n$. He first generates $n$ shares of the secret $S$, denoted by $S_1,S_2,\cdots,S_n$, and then distributes $S_i$ to $P_i$ for $1\leq i\leq n$ such that any $t$ or more participants are able to recover the secret $S$ together (say the recovery property) and any $t-1$ or less participants obtain no information about the secret $S$ (called the perfectly secure property). It is easy to see that this $(t,n)$ threshold scheme is highly reliable and secure for the secret $S$, since it can resist up to $n-t$ erasure errors or the leakage of up to $t-1$ shares. McEliece and Sarwate \cite{McEliece-Sarwate-1981} indicated the close relationship between Shamir's $(t,n)$ threshold scheme and $[n,t]$ Reed-Solomon codes \cite{M-S}. Lai and Ding \cite{Lai-Ding-2004} proposed several generalizations of Shamir's threshold scheme in 2004. Recently, new generalizations of threshold secret sharing schemes by using integer polymatroids were proposed by Farr\`{a}s \textit{et al.} \cite{Xingchaoping-2014}. Benaloh and Leichter \cite{JJ-1988} introduced the ``weighted secret sharing scheme'' to assign multiple shares rather than a single share to some participants. In \cite{Stinson-2010}, Nojoumian \textit{et al.} introduced the notion of social secret sharing scheme to renew shares of a participant based on his ``reputation''. Furthermore, for the``one time'' problem of lots of secret sharing schemes, meaning that the secret and shares are known to everyone after a public secret recovery process, Nojoumian and Stinson \cite{Stinson-2013} provided threshold schemes to generate new secrets dynamically in the absence of the dealer.

In addition, for practical applications, sometimes the share of a participant will fail such as destroyed or lost. So how to recover the failed share in order to keep the stability and the robustness of the system is meaningful and interesting. One natural approach is to request the dealer to regenerate and redistribute an appropriate share to the corresponding participant whose original share fails. But it is not difficult to see that this approach has a plenty of shortcomings. For instances, it has to seek the help of the dealer, and the dealer must maintain the original secret and the regenerating and redistributing functions to obtain a new share for the corresponding participant. Moreover, it possibly costs a lot to request the service of the dealer in practice. Thus, in order to avoid these shortcomings, we are eager to construct a new secret sharing scheme with repairing function, where the failed share can be repaired only by some other participants without the help of the dealer, and meanwhile, the properties of recovery and perfect security of the system are still guaranteed.

Motivated by the goal as mentioned above, the recent research on regenerating codes in distributed storage systems \cite{Dimakis-etc-2010} may be feasible to solve our problem since regenerating codes have a better repairing property. To be specific, in a distributed storage system of $n$ storage nodes $P_1,P_2,\cdots,P_n$, let $M$ be the original data, regarded as an $|M|$-dimensional row vector. An $(n,k,d)$ regenerating code with parameters $(\alpha,\beta)$ and an $|M|\times n\alpha$ generator matrix $G=[G_1,G_2,\cdots,G_n]$ is used to distribute the original data $M$ to the $n$ storage nodes, where each $G_i$, $1\leq i\leq n$, is an $|M|\times \alpha$ submatrix of $G$. Subsequently, each storage node $P_i$ stores $\alpha$ symbols by computing $M\cdot G_i$ for $1\leq i\leq n$. This regenerating code should satisfy the requirements that any $k$ nodes can recover the original data $M$ and any $d$ intact nodes can repair a failed node by providing $\beta$ symbols (encoded usually) from each intact one. In particular, the parameters of $(n,k,d)$ regenerating codes satisfy the equation $|M|=\sum_{i=0}^{k-1} {\rm min}\{\alpha,(d-i)\beta\}$. Furthermore, Dimakis \textit{et al.} in \cite{Dimakis-etc-2010} defined two special classes of regenerating codes, the minimum bandwidth regenerating (MBR) codes satisfying $\alpha=d\beta$ which guarantee the minimum of the total transmitted symbols for repair, and the minimum storage regenerating (MSR) codes satisfying $\alpha=(d-k+1)\beta$ which ensure the minimum of the stored symbols for each node. More works about MBR codes and MSR codes can be referred to \cite{Rashmi-Shah-Kumar-2011} and \cite{Lin-Chung-2014}.

Although the goals of secret sharing schemes and distributed storage systems are totally different, one for sharing the secret and the other for storing the data, the requirements of recovering the secret and the original data are similar. In addition, the repairing property of regenerating codes may be helpful to accomplish the repairing function of a secret sharing scheme. Therefore, we naturally try to combine a regenerating code and a threshold secret sharing scheme to obtain a secret sharing schemes with repairing function. Review the construction of Shamir's $(t,n)$ threshold scheme, which firstly selects at random a polynomial $f(x)=\sum^{t-1}_{i=0}a_ix^i$ of degree $t-1$ over a finite field $\Fq$, where $a_0=s$ is considered as the secret, and then generates $n$ evaluations $s_i=f(x_i)$, $1\leq i\leq n$, for $n$ distinct participants. Our initial and natural idea is to concatenate the Shamir's threshold secret sharing scheme and a regenerating code directly. More specifically, we first use $f(x)=s+\sum^{t-1}_{i=1}a_ix^i$ to generate $t$ distinct evaluations, which together can recover the secret $s$. Then regard the $t$ evaluations as original data in distributed systems and use an $(n,k,d)$ regenerating code to obtain $n$ distinct shares for all $n$ participants. Clearly, the regenerating code can ensure the repair of any failed share, and any $k$ participants can recover all $t$ evaluations, further recover the secret $s$. However, it is very unfortunate that the fundamental perfectly secure property is no more satisfied for most cases. Particularly, a specific example will be given to show this point in the next section. Thus, we have to reconsider this problem. In this paper, we propose an innovative construction of repairable threshold secret sharing schemes by applying linearized polynomials as an important component, which can qualify all desired properties, particularly, perfectly secure property.

The remainder of the paper is organized as follows. In Section \ref{Construction}, we formally define repairable threshold secret sharing schemes and propose a feasible construction. Further, the functions, including recovery function and repairing function, and perfectly secure property of the proposed construction are analyzed in detail. In Section \ref{Evaluation}, in order to analyze the performance of repairable threshold secret sharing schemes, we introduce a new repairing rate to characterize the efficiency of repairing function and derive its tight upper bound. Subsequently, we also discuss the traditional information rate when the optimal repairing rate is achieved, which describes the efficiency of the storage for each participant. Furthermore, we show that the repairable threshold schemes obtained by our construction can achieve both types of the optimality by applying MBR codes. Finally, the paper is concluded in Section \ref{Conclusion}.

\section{Repairable Threshold Secret Sharing Schemes}\label{Construction}
\label{sec:1}
In this section, we will discuss repairable threshold secret sharing schemes in detail. As mentioned in the above section, our initial and natural idea is to directly combine Shamir's threshold scheme and a regenerating code to obtain a repairable threshold secret sharing scheme. Specifically, for a given secret $s$, we first select a polynomial $f(x)=s+\sum^{t-1}_{i=1}a_ix^i$ of degree $t-1$ at random over a finite field $\Fq$, and then we compute $t$ distinct evaluations of $f(x)$, denoted by $y_1,y_2,\cdots,y_t$. After that, use an $(n,k,d)$ regenerating code with generator matrix $G=[G_1,G_2,\cdots,G_n]$ to distribute these $t$ evaluations. Hence, the shares are obtained by computing $s_i=[y_1,\cdots,y_t]\cdot G_i$, $1\leq i\leq n$, for all $n$ participants. Evidently, this scheme satisfies the following: (i) any $k$ or more participants can recover the $t$ evaluations, and further recover the secret $s$ by polynomial interpolation. (ii) any failed share can be repaired successfully by any other $d$  participants because of the repairing property of the regenerating code. But unfortunately, the necessary perfectly secure property is no longer satisfied as what the following example illustrates.

\begin{eg}\label{eg}
A secret $s$ in the finite field $\mathbb{F}_{11}$ needs to be shared among $4$ participants $P_1,P_2,P_3,P_4$. In the following, we give a $(t=2,n=4)$ threshold secret sharing scheme by combining Shamir's threshold scheme and an $(n=4,k=2,d=3)$ regenerating code directly. At first, randomly select $3$ elements $a_1,a_2,a_3$ in $\mathbb{F}_{11}$ independently and uniformly. Then define a polynomial: $$f(x)=s+\sum\limits_{j=1}^3 a_j\cdot x^j\in \mathbb{F}_{11}[x].$$
Then, calculate $y_i=f(x_i)$, where $x_i=i$, $i=1,2,3,4\in \mathbb{F}_{11}$, and all of $x_i$ are public. Use a $(4,2,3)$ regenerating code with the public generator matrix
$$G=[G_1|G_2|G_3|G_4]=\left[\begin{array}{cc|cc|cc|cc}
1 & 0 & 0 & 0 & 0 & 5 & 1 & 1\\ 0 & 1 & 0 & 0 & 2 & 1 & 9 & 0\\
0 & 0 & 1 & 0 & 1 & 1 & 2 & 4\\ 0 & 0 & 0 & 1 & 1 & 3 & 0 & 4
\end{array}\right]$$
to generate the share $s_i$ of participant $P_i$ by computing $$s_i=[y_1,y_2,y_3,y_4]\cdot G_i$$ for $1\leq i\leq 4$. Thus, we obtain
\begin{align*}
s_1&=[y_1,y_2],\ s_2=[y_3,y_4],\\
s_3&=[2y_2+y_3+y_4,5y_1+y_2+y_3+3y_4],\\
s_4&=[y_1+9y_2+2y_3,y_1+4y_3+4y_4].
\end{align*}

By the knowledge of regenerating codes, it is apparent to verify that any $2$ or more participants can recover the secret $s$ and any failed share can be repaired by any other $3$ shares. But this threshold scheme isn't perfectly secure for the secret $s$ as analyzed below. Take $P_3$ for example, whose share $s_3$ is $[2y_2+y_3+y_4,5y_1+y_2+y_3+3y_4]$. Substituting $y_i=s+\sum_{j=1}^3 a_j\cdot i^j$, $i=1,2,3,4\in \mathbb{F}_{11}$, we have the following two linear equations:
\begin{align*}
\begin{cases}
2y_2+y_3+y_4=4s+8a_3,\\
5y_1+y_2+y_3+3y_4=10s+a_3.
\end{cases}
\end{align*}
Thus, the participant $P_3$ himself can recover the secret $s$ by solving the above two linear equations. This conflicts to the perfectly secure requirement, that is, any one participant can obtain nothing about the secret $s$ in the information-theoretical sense.
\end{eg}

In fact, due to the additional repairing requirement, the shares between different participants have some correlations, which leads to the failure of perfect security essentially. Since the direct idea cannot achieve all goals, some other techniques have to be considered. Thus, we further introduce linearized polynomials as a component to construct repairable secret sharing schemes, because linear operations are preserved for linearized polynomials to be interpreted below. First, we give the definition of $(n,k,d)$ repairable threshold secret sharing scheme formally as follows.

\begin{defn}\label{defn_1}
A threshold secret sharing scheme (TSSS) is called an $(n,k,d)$ repairable threshold secret sharing scheme (repairable-TSSS), if the following three conditions are qualified:
\begin{enumerate}
  \item any $k$ or more out of $n$ participants can recover the secret;
  \item any $k-1$ or less participants obtain no information about the secret;
  \item any $d$ ($\geq k$\footnote{It is natural, as otherwise $d<k$ will lead to a contradiction to the condition 1) in Definition \ref{defn_1}.}) intact participants can repair any another share of some participant provided that it fails.
\end{enumerate}
\end{defn}

\begin{rem}
In this paper, the failed share can be repaired by any other $d$ shares, and we always assume that each of these $d$ participants provides $\beta$ $(<\alpha)$ symbols for the repair process, where $\alpha$ represents the size of the share. This implies that, in order to repair a failed share, each participant just provides a piece of his share for the repair, which is necessary and reasonable in sense of security between shares since no participant is willing to contribute all of his own share to repair a failed one. In fact, we further require that all transmitted $d$ pieces of shares together, of size $d\beta$ totally, still obtain nothing about the secret.
\end{rem}

Next, we will illustrate our construction of $(n,k,d)$ repairable-TSSSs explicitly below.

\noindent \textbf{Construction:}
\begin{description}
  \item[Step 1:]\ let $t=\sum\limits_{i=0}^{k-1}\min\{ \alpha,\ (d-i)\beta\}$, and then select randomly $t-1$ elements $a_i$, $1\leq i \leq t-1$, in $\mathbb{F}_{p^m}$ independently and uniformly. Then define a linearized polynomial as follows:
       $$f(x)=sx+\sum\limits_{i=1}^{t-1} a_i x^{p^i}\in \mathbb{F}_{p^m}[x].$$
  \item[Step 2:]\ select $t$ distinct elements $x_i$ in $\mathbb{F}_{p^m}$, $1\leq i \leq t$, such that they are linearly independent over the prime field $\mathbb{F}_{p}$. So it is necessary to require $m\geq t$. Then publish $x_i$ and compute the evaluation $y_i=f(x_i)$ for $1\leq i \leq t$.
  \item[Step 3:]\ let $\mathcal{C}$ be an $(n,k,d)$ regenerating code with $\mathbb{F}_p$-valued generator matrix $G=[G_1,G_2,\cdots,G_n]$, where each $G_i$, $1\leq i \leq n$, is a $t\times \alpha$ full column rank matrix, i.e., $\Rank(G_i)=\alpha$. Each share $s_i$ is generated by computing $s_i=[y_1,y_2,\cdots,y_t]\cdot G_i$, which is an $\alpha$-dimensional $\mathbb{F}_{p^m}$-valued vector. Subsequently, distribute $s_i$ to $P_i$ for $1\leq i \leq n$. Notice that $\mathcal{C}$ satisfies the repair function that $d$ intact shares can repair a failed one by providing $\beta$ symbols for each one.
\end{description}

\subsection{Function Analysis}\label{sec:2}

In the following, we will analyze the functions of recovery of the secret and repair of the failed share. We will need the following important property before discussion further.

\begin{lem}[{\cite[Lemma 3.51]{R-H-P-1997}}]\label{invertible of coefficient matrix}
Let $x_1,x_2,\cdots,x_n \in\mathbb {F}_{p^m},$ $(m\geq n)$, then
$$\det\left[\begin{matrix}x_1&x_1^p&x_1^{p^2}&\cdots&x_1^{p^{n-1}}\\x_2&x_2^p&x_2^{p^2}&\cdots&x_2^{p^{n-1}}\\
\cdots&\cdots&\cdots&\cdots&\cdots\\x_n&x_n^p&x_n^{p^2}&\cdots&x_n^{p^{n-1}}\end{matrix}\right]\neq 0$$
if and only if $x_1,x_2,\cdots,x_n$ are linearly independent over the prime field $\mathbb {F}_p$.
\end{lem}

\begin{thm}\label{thm_recover}
For the constructed $(n,k,d)$ repairable-TSSSs, any $k$ participants can recover the secret $s$ together, and any $d$ intact participants, each of which provides $\beta$ symbols, can repair a failed share.
\end{thm}
\begin{IEEEproof}
Since $\mathcal{C}$ is an $(n,k,d)$ regenerating code, any $k$ participants can recover the $t$ evaluations $y_1, y_2, \cdots, y_t$. That is, they obtain the following $t$ equations of $t$ variables $s,a_1,\cdots,a_{t-1}$:
$$sx_{j}+\sum_{i=1}^{t-1}a_i x_{j}^{p^i}=y_{j},\ 1\leq j \leq t.$$
Since $x_1,x_2,\cdots,x_t$ are linearly independent over the prime field $\mathbb{F}_p$, together with Lemma \ref{invertible of coefficient matrix}, the ${t\times t}$ coefficient matrix
\begin{align}\label{coefficient-matrix}
X=\begin{bmatrix}
x_1 & x_2 & \cdots & x_t\\
x_1^p & x_2^p & \cdots & x_t^p\\
\cdots & \cdots & \cdots & \cdots \\
x_1^{p^{t-1}} & x_2^{p^{t-1}} & \cdots & x_t^{p^{t-1}}
\end{bmatrix}
\end{align}
with respect to the above $t$ linear equations is invertible. Thus, this linear system has a unique solution, that is, recovers $s$. Moreover, the repairing function of this TSSS is evident as the repairing property of the used $(n,k,d)$ regenerating code $\mathcal{C}$. we accomplish the proof.
\end{IEEEproof}

\subsection{Security Analysis}
\label{sec:3}
Next, we will show that our proposed repairable-TSSSs are also perfectly secure, that is, any $k-1$ or less participants obtain nothing about the secret. First, we need the following dimension lemma on regenerating codes. Again, let $M$ be the original data, regarded as a $t$-dimensional row vector. We consider an $(n,k,d)$ regenerating code with the $t\times n\alpha$ generator matrix $G=[G_1,G_2,\cdots,G_n]$, where each $G_i$ is a $t \times \alpha$ submatrix. Each node $P_i$ stores $\alpha$ linear combinations of the original data $M$ by computing $M G_i$ for $1\leq i \leq n$. Let $W_i$ denote the vector space spanned by the $\alpha$ column vectors of $G_i$, $1\leq i \leq n$, and clearly $\dim(W_i)\leq \alpha$. When a node fails, any other $d$ intact nodes can repair it by providing $\beta$ symbols for each one. Thus, it also follows $\dim(W_i)\leq d \beta$.

In order to show the perfectly secure property, we need the following dimension lemma.
\begin{lem}\label{dimension-lemma}
For any $(n,k,d)$ regenerating code with parameters $(\alpha,\beta)$, let $W_{i_1}$, $W_{i_2}$, $\cdots$, $W_{i_{k-1}}$ be $k-1$ vector spaces corresponding to any $k-1$ nodes $P_{i_1}, P_{i_2}, \cdots, P_{i_{k-1}}$. Then the dimension of the summation of these $k-1$ vector spaces is strictly smaller than $t$, i.e., the size of the original data $M$.
\end{lem}
\begin{IEEEproof}
Consider any $k-1$ nodes. Without loss of generality, assume that they are the first $k-1$ nodes $P_1,P_2,\cdots,P_{k-1}$, and their corresponding vector spaces are $W_1,W_2,\cdots,W_{k-1}$ of dimensions $\Omega_{1}, \Omega_{2}, \cdots, \Omega_{k-1}$, respectively. By using the dimension theorem recursively, one has
\begin{align}\label{eq:a1}
&\dim\Big(\sum_{i=1}^{k-1}{W_{i}}\Big)=\sum_{i=1}^{k-1}\dim\Big(W_{i}\Big)-\sum_{j=1}^{k-2}\dim\Big(W_j \cap \sum_{i=j+1}^{k-1}W_{i}\Big).
\end{align}

Further, notice that for any $j, 1\leq j\leq k-2$, if the node $P_j$ fails, it can be repaired by $P_{j+1}$, $P_{j+2}$, $\cdots$, $P_{k-1}$ and the other $d-(k-1-j)$ nodes not including himself $P_j$. The latter $d-(k-1-j)$ nodes provide at most $(d-(k-1-j))\beta$ dimensions for the vector space $W_j$ corresponding to the failed node $P_j$. Hence all remaining dimensions of $W_j$ must come from the nodes $P_{j+1},P_{j+2},\cdots,P_{k-1}$, i.e.,
\begin{align*}
\dim\Big(W_j\cap\sum_{i=j+1}^{k-1}W_{i}\Big)\geq \max\{0,\Omega_j-(d-(k-1-j))\beta\}.
\end{align*}

Combining with the equality (\ref{eq:a1}), we further have
\begin{align*}
&\dim\Big(\sum_{i=1}^{k-1}W_{i}\Big) \\
\leq&  \sum_{i=1}^{k-1}\Omega_i-\sum_{j=1}^{k-2}\max\{0, \Omega_j-(d-(k-1-j))\beta\}\\
=&\Omega_{k-1}+\sum_{j=1}^{k-2} [\Omega_j-\max\{0, \Omega_j-(d-(k-1-j))\beta\}]\\
=&\Omega_{k-1}+\sum_{j=1}^{k-2} \min\{ \Omega_j, (d-(k-1-j))\beta\}\\
\leq& \min\{\alpha,d\beta \}+\sum_{j=1}^{k-2}\min\{\alpha,(d-(k-1-j))\beta\}\\
=&\sum_{j=1}^{k-1}\min\{\alpha, (d-(k-1-j))\beta\}\\
=&\sum_{i=0}^{k-2}\min\{\alpha, (d-i)\beta\}\\
<&\sum_{i=0}^{k-1}\min\{\alpha, (d-i)\beta\}=t,
\end{align*}
where the last inequality follows from $\min\{\alpha, (d-(k-1))\beta\}>0$. This completes the proof.
\end{IEEEproof}

\begin{thm}\label{secure-thm}
The constructed $(n,k,d)$ repairable-TSSSs are perfectly secure.
\end{thm}
\begin{IEEEproof}
For the perfectly secure property, we indicate that any $k-1$ or less participants cannot obtain any information about the secret $s$. By Lemma \ref{dimension-lemma} and our construction, any $k-1$ or less participants correspond to a vector space of dimension less than $t$, say dimension $r$ and $r<t$. Let $r$ $t$-column vectors $\vec{g}_1, \vec{g}_2, \cdots, \vec{g}_r$ be arbitrary basis of this vector space, and further
$$\hat{G}=\big[ \vec{g}_1, \vec{g}_2, \cdots, \vec{g}_r \big]\triangleq \left[\begin{matrix}
g_{1,1} & g_{1,2} & \cdots & g_{1,r}\\
g_{2,1} & g_{2,2} & \cdots & s_{2,r}\\
\cdots  & \cdots  & \cdots & \cdots\\
g_{t,1} & g_{t,2} & \cdots & g_{t,r}
\end{matrix}\right],$$
which is a $t\times r$ full column rank matrix over the prime field $\mathbb{F}_p$. Subsequently, let $z_1,z_2,\cdots, z_r$ be $r$ linear combinations of $t$ evaluations $y_1,y_2,\cdots,y_t$, satisfying
$$[z_1,z_2,\cdots, z_r]=[y_1,y_2,\cdots, y_t]\cdot\hat{G}.$$
Further, since
$$[y_1,y_2,\cdots, y_t]=[s,a_1,\cdots,a_{t-1}]\cdot X,$$ where $X$ is a $t\times t$ invertible matrix over $\mathbb{F}_{p^m}$ as defined in (\ref{coefficient-matrix}), we deduce
\begin{align*}
&[z_1,z_2,\cdots, z_r]=[y_1,y_2,\cdots, y_t]\cdot\hat{G}\\
=&[s,a_1,\cdots,a_{t-1}]\cdot X\hat{G}
\end{align*}
\begin{align*}
=&[s,a_1,\cdots,a_{t-1}]
\left[\begin{matrix}
x_1 & x_2 & \cdots & x_t\\
x_1^p & x_2^p & \cdots & x_t^p\\
\cdots & \cdots & \cdots & \cdots \\
x_1^{p^{t-1}} & x_2^{p^{t-1}} & \cdots & x_t^{p^{t-1}}
\end{matrix}\right]_{t\times t}\cdot
\left[\begin{matrix}
g_{1,1} & g_{1,2} & \cdots & g_{1,r}\\
g_{2,1} & g_{2,2} & \cdots & s_{2,r}\\
\cdots  & \cdots  & \cdots & \cdots\\
g_{t,1} & g_{t,2} & \cdots & g_{t,r}
\end{matrix}\right]_{t\times r}\\
=&[s,a_1,\cdots,a_{t-1}]\cdot \left[\begin{matrix}
\sum\limits_{i=1}^t g_{i,1}x_i & \sum\limits_{i=1}^t g_{i,2}x_i & \cdots & \sum\limits_{i=1}^t g_{i,r}x_i\\
\sum\limits_{i=1}^t g_{i,1}x_i^p & \sum\limits_{i=1}^t g_{i,2}x_i^p & \cdots & \sum\limits_{i=1}^t g_{i,r}x_i^p\\
\cdots  & \cdots  & \cdots & \cdots\\
\sum\limits_{i=1}^t g_{i,1}x_i^{p^{t-1}} & \sum\limits_{i=1}^t g_{i,2}x_i^{p^{t-1}} & \cdots & \sum\limits_{i=1}^t g_{i,r}x_i^{p^{t-1}}
\end{matrix}\right]_{t\times r}\\
=&[s,a_1,\cdots,a_{t-1}]\cdot
\left[\begin{matrix}
\sum\limits_{i=1}^t g_{i,1}x_i & \sum\limits_{i=1}^t g_{i,2}x_i & \cdots & \sum\limits_{i=1}^t g_{i,r}x_i\\
(\sum\limits_{i=1}^t g_{i,1}x_i)^p & (\sum\limits_{i=1}^t g_{i,2}x_i)^p & \cdots & (\sum\limits_{i=1}^t g_{i,r}x_i)^p\\
\cdots  & \cdots  & \cdots & \cdots\\
(\sum\limits_{i=1}^t g_{i,1}x_i)^{p^{t-1}} & (\sum\limits_{i=1}^t g_{i,2}x_i)^{p^{t-1}} & \cdots & (\sum\limits_{i=1}^t g_{i,r}x_i)^{p^{t-1}}
\end{matrix}\right].
\end{align*}

Since $\hat{G}$ is full column rank, and $x_i$, $1\leq i \leq t$, are linearly independent over $\mathbb{F}_p$, it follows that $\sum_{i=1}^t g_{i,j}x_i$, $1\leq j \leq r$, are also linearly independent over $\mathbb{F}_p$. Together with Lemma \ref{invertible of coefficient matrix}, it is implied that $X\hat{G}$ is still full column rank. Further, we define another matrix
$$\hat{X}=\left[\begin{matrix}
x_1^p & x_2^p & \cdots & x_t^p\\
x_1^{p^{2}} & x_2^{p^{2}} & \cdots & x_t^{p^{2}}\\
\cdots & \cdots & \cdots & \cdots \\
x_1^{p^{t-1}} & x_2^{p^{t-1}} & \cdots & x_t^{p^{t-1}}
\end{matrix}\right],$$
which is a submatrix of $X$ by deleting the first row of $X$, and then calculate $\hat{X}\hat{G}$ of size $(t-1)\times r$ below:
\begin{align*}
&\hat{X}\hat{G}=\left[\begin{matrix}
(\sum\limits_{i=1}^t g_{i,1}x_i)^p & (\sum\limits_{i=1}^t g_{i,2}x_i)^p & \cdots & (\sum\limits_{i=1}^t g_{i,r}x_i)^p\\
(\sum\limits_{i=1}^t g_{i,1}x_i)^{p^2} & (\sum\limits_{i=1}^t g_{i,2}x_i)^{p^2} & \cdots & (\sum\limits_{i=1}^t g_{i,r}x_i)^{p^2}\\
\cdots  & \cdots  & \cdots & \cdots\\
(\sum\limits_{i=1}^t g_{i,1}x_i)^{p^{t-1}} & (\sum\limits_{i=1}^t g_{i,2}x_i)^{p^{t-1}} & \cdots & (\sum\limits_{i=1}^t g_{i,r}x_i)^{p^{t-1}}
\end{matrix}\right].
\end{align*}
By Lemma \ref{invertible of coefficient matrix}, this matrix $\hat{X}\hat{G}$ is also full column rank since $\Big(\sum_{i=1}^t g_{i,j}x_i\Big)^p$, $1\leq j \leq r$, are also linearly independent over $\mathbb{F}_p$.

Next, we will prove that $s$ is perfectly secure by deducing ${\rm{H}}(S)={\rm{H}}(S|Z)$, where $S$ represents the corresponding random variable of the secret $s$, and $Z=[Z_1,Z_2,\cdots,Z_r]$ denotes the corresponding random vector of the above $r$-row vector $[z_1,z_2,\cdots, z_r]$. Herein, ${\rm{H}}(\cdot)$ and ${\rm{H}}(\cdot|\cdot)$ represent entropy and conditional entropy, respectively. Thus, in order to show ${\rm{H}}(S)={\rm{H}}(S|Z)$, it suffices to prove
$$P_r(S=s)=P_r(S=s|Z=[z_1,z_2,\cdots, z_r])$$
for all possible $s$ and $z_i$, $1\leq i\leq r$, in $\mathbb{F}_{p^m}$.

Subsequently, due to Bayes formula
$$p(s|{z})=\frac{p(s,{z})}{p({z})}=\frac{p({z}|s)p(s)}{p({z})},$$
it suffices to prove $$P_r({Z=[z_1,z_2,\cdots, z_r]})=P_r({Z=[z_1,z_2,\cdots, z_r]}|S=s).$$
Evidently, the event ``$Z=[z_1,z_2,\cdots,z_r]$'' is equivalent to the event ``$S=s,A_i=a_i,1\leq i\leq t-1$, such that $[z_1,z_2,\cdots, z_r]=[s,a_1,\cdots,a_{t-1}]\cdot X\hat{G}$'', where $A_i$, $1\leq i\leq t-1$, is the corresponding random variable of the coefficient $a_i$ of $x^{p^i}$ in the used linearized polynomial $f(x)=sx+\sum_{i=1}^{t-1} a_i x^{p^i}\in \mathbb{F}_{p^m}[x]$, all of which are independently and uniformly distributed over $\mathbb{F}_{p^m}$. Thus, one obtains
\begin{align}
&P_r\Big({Z}=[z_1,z_2,\cdots,z_r]\Big) \notag\\
=&P_r\Big([S,A_1,\cdots,A_{t-1}]=[s,a_1,\cdots,a_{t-1}]\ {\rm such\ that} \notag\\
&\qquad~~\qquad\qquad[s,a_1,\cdots,a_{t-1}]\cdot X\hat{G}=[z_1,z_2,\cdots, z_r]\Big) \notag\\
=&\frac{(p^m)^{t-r}}{(p^m)^t}=\frac{1}{(p^m)^{r}}, \label{eq}
\end{align}
where the first equality in (\ref{eq}) follows since the $t\times r$ matrix $X\hat{G}$ is full column rank. Similarly,
\begin{align}
&P_r\Big({Z}=[z_1,z_2,\cdots,z_r]|S=s\Big) \notag\\
=&P_r\Big([A_1,A_2,\cdots,A_{t-1}]=[a_1,a_2,\cdots,a_{t-1}]\ {\rm such\ that} \notag\\
& [a_1,a_2,\cdots,a_{t-1}]\cdot \hat{X}\hat{G}=[z_1-s\sum\limits_{i=1}^t g_{i,1}x_i,\cdots, z_r-s\sum\limits_{i=1}^t g_{i,r}x_i]\Big) \notag\\
=&\frac{(p^m)^{t-r-1}}{(p^m)^{t-1}}=\frac{1}{(p^m)^{r}}, \label{eqq}
\end{align}
where similarly the first equality in (\ref{eqq}) follows since the $(t-1)\times r$ matrix $\hat{X}\hat{G}$ is full column rank.
Thus, we further deduce ${\rm{H}}(S)={\rm{H}}(S|{Z})$, and notice that ${\rm{H}}(S)={\rm{H}}(S|{Z})$ follows for all such invertible matrices $\hat{G}$. This shows that any $k-1$ participants can obtain no information about the secret, which indicates the perfectly secure property satisfied. This completes the proof.
\end{IEEEproof}

\section{Performance Analysis of Repairable-TSSSs}\label{Evaluation}
In this section, we focus on the performance analysis of the proposed repairable-TSSSs. Since repairable-TSSSs contain an additional function of repair, we define a repairing rate below to characterize the efficiency of repair, which is also important in the aspect of security. For an $(n,k,d)$ repairable-TSSS, if one share fails, say $s_i$, any other $d$ intact participants can repair it. We further use $T$ to denote the index set of the $d$ participants, say a \emph{repairing set}. Further, for each $j\in T$, let $R_{j,T}^i$ represent the symbols provided by $P_j$ to repair the failed share $s_i$. Now, we can give the definition of the repairing rate.

\begin{defn}
For a repairable-TSSS, the repairing rate of every participant $P_i$, $1\leq i \leq n$, is defined as the ratio
$$\rho^{(i)}_{\rm rep}={\rm min}\{\frac{ {\rm log}_2|\mathcal {S}_i|}{{\rm log}_2\prod\limits_{j\in T}|\mathcal {R}_{j,T}^i|},T\subseteq[n]\backslash\{i\}\ {\rm with}\ |T|=d\},$$
where $[n]=\{1,2,\cdots,n\}$, and $\mathcal {S}_i$ and $\mathcal {R}_{j,T}^i$ represent the sets of all possible values taken by $s_i$ and $R_{j,T}^i$, respectively. Further, the repairing rate of this scheme is defined as $$\rho_{\rm rep}={\mathop{\rm min}\limits_{1\leq i\leq n}}\rho^{(i)}_{\rm rep}.$$
\end{defn}

Actually, $\rho^{(i)}_{\rm rep}$ is the minimum of the ratios of the number of bits in the share $s_i$ to the total number of bits provided by all $d$ participants in $T$ for repairing $s_i$ amongst all sets $T\subseteq [n]\backslash\{i\}$ of size $d$. This naturally characterizes the efficiency of the repair of $P_i$. In addition, this repairing rate also describes the security of the repair, since $P_i$ possibly obtains some extra information about other participants' shares during the repairing process. The smaller the $\rho^{(i)}_{\rm rep}$ is, the more extra information $P_i$ can obtain. Thus, higher repairing rate $\rho^{(i)}_{\rm rep}$ is preferable, so is the repairing rate $\rho_{\rm rep}$. For the repairable-TSSSs, all shares $s_i$, $1\leq i\leq n$,  are $\alpha$-dimensional vectors and each $R_{j,T}^i$, $j\in T\subseteq [n]$, can be regarded as a $\beta$-dimensional vector over $\mathbb{F}_{p^m}$, and thus $|\mathcal {S}_i|=p^{m\alpha}$ and $|\mathcal {R}_{j,T}^i|=p^{m\beta}$. Consequently, the repairing rate of the threshold-TSSSs is $\alpha/{d\beta}$. Note that $\alpha\leq d\beta$ is necessary for successful repair. Hence, the repairing rate $\rho_{\rm rep}$ of the scheme is upper bounded by $1$, i.e., $\rho_{\rm rep}\leq 1$. We say it optimal repairing rate when this upper bound is achieved. Actually, besides achieving the maximum efficiency of repair, a repairable-TSSS with optimal repairing rate is more secure. Specifically, if $\rho_{\rm rep}<1$, that is, $\alpha< d\beta$, the participant of failed share possibly obtains extra information about other participants' shares in addition to the necessary information for the repair. Apparently, this case is not secure enough, since a malicious participant can always claim his share fails in order to obtain extra information, and after enough repairing processes, he even possibly obtains enough information to recover the secret $s$. Thus, we always expect to use repairable-TSSSs with optimal repairing rate.
Furthermore, in secret sharing theory, the \emph{information rate}, as a traditional index to characterize the efficiency of the storage performance of secret sharing schemes, is defined below.

\begin{defn}[{\cite[Definition 13.4]{Stinson-2006}}]
For a secret sharing scheme, the information rate for every participant $P_i$, $1\leq i \leq n$, is defined as the ratio:
$$\rho^{(i)}_{\rm inf}=\frac{ {\rm log}_2|\mathcal {S}|}{{\rm log}_2|\mathcal {S}_i|},$$
where $\mathcal {S}$ and $\mathcal {S}_i$ are the sets of all possible values taken by $s$ and $s_i$, respectively. Further, the information rate of this scheme is defined as:
$$\rho_{\rm inf}={\mathop{\rm min}\limits_{1\leq i\leq n}}\rho^{(i)}_{\rm inf}.$$
\end{defn}

It is not difficult to see that this rate characterizes the storage efficiency of TSSSs. Next, we discuss the optimal information rate of the repairable-TSSSs under the optimal repairing rate.

\begin{thm}\label{thm-inf-rate}
For the proposed $(n,k,d)$ repairable-TSSSs with optimal repairing rate, i.e., $\rho_{\rm rep}=1$, the information rate $\rho_{\rm inf}$ is upper bounded by $\frac{k(2d-k+1)}{2dt}$.
\end{thm}
\begin{IEEEproof}
Reviewing our proposed threshold scheme, for a given $t$, let $S$ and $A_1,A_2,\cdots,A_{t-1}$ be $t$ random variables corresponding all coefficients of the used linearized polynomial over $\mathbb{F}_{p^m}$, all of which are chosen independently and uniformly over $\mathbb{F}_{p^m}$. Each share is an $\alpha$-dimensional vector over $\mathbb{F}_{p^m}$ and the data provided by a participant for repair is regarded as a $\beta$-dimensional vector over $\mathbb{F}_{p^m}$. Notice that the information rate of our scheme is $1/\alpha$ evidently since $|\mathcal {S}|=p^{m}$ and $|\mathcal {S}_i|=p^{m\alpha}$, $1\leq i\leq n$. In the following, it suffices to discuss the lower bound on $\beta$, because $\alpha=d\beta$ under optimal repairing rate.

Consider arbitrary $k$ participants. Without loss of generality, assume that they are $P_1,P_2,\cdots,P_k$ and those random variables representing their shares are $S_1,S_2,{\cdots},S_k$, respectively. Then, it follows
\begin{align}
t{\rm{H}}(S)  &= {\rm{H}}(S,A_1,\cdots,A_{t-1})= {\rm{H}}(S_l,1\leq l\leq k)\notag\\
&\leq {\rm{H}}(R_{u,T}^1, u\in T\backslash\{2,\cdots,k\}, R_{v,T}^1, v\in \{2,\cdots,k\}, \notag\\
& \qquad   S_l, 2\leq l\leq k) \label{eq:a2}\\
&= {\rm{H}}(R_{u,T}^1, u\in T\backslash\{2,\cdots,k\}, S_l, 2\leq l\leq k) \label{eq:a3}\\
&\leq {\rm{H}}(R_{u,T}^1, u\in T\backslash\{2,\cdots,k\})+{\rm{H}}(S_l,2\leq l\leq k)\notag\\
&\leq (d-(k-1))\beta {\log_2}\ {p^m}+{\rm{H}}(S_l,2\leq l\leq k) \label{eq:a4}\\
& = (d-k+1)\beta {\rm{H}}(S)+{\rm{H}}(S_l,2\leq l\leq k),\notag
\end{align}
where $T$ in $(\ref{eq:a2})$ is a repairing set of $S_1$, containing the indices of $P_{2}$, $P_{3},{\small\cdots},P_k$ and other $d-(k-1)$ intact participants, and thus the inequality $(\ref{eq:a2})$ follows because $R_{u,T}^1$, $u\in T$, can repair $S_1$; the equality $(\ref{eq:a3})$ follows from the fact ${\rm{H}}(R^1_{l,T}|S_l)=0$ for $2\leq l\leq k$; and the inequality $(\ref{eq:a4})$ follows because each of $R_{u,T}^1$, $u\in T\backslash\{2,\cdots,k\}$, is regarded as a $\beta$-dimensional random vector taking values in $\mathbb{F}_{p^m}^{\beta}$. Applying the same analysis method on ${\rm{H}}(S_l,2\leq l\leq k)$, we further obtain
\begin{align*}
 {\rm{H}}(S_l,2\leq l\leq k)\leq (d-k+2)\beta {\rm{H}}(S)+{\rm{H}}(S_l,3\leq l\leq k).
\end{align*}
So far and so forth, we finally obtain
\begin{align*}
t{\rm{H}}(S)\leq\sum\limits^{k-1}_{i=0}(d-i)\beta {\rm{H}}(S)=\frac{k(2d-k+1)}{2}\beta {\rm{H}}(S),
\end{align*}
which implies $\beta\geq \frac{2t}{k(2d-k+1)}$, further indicating
$$\rho_{\rm inf}=\frac{ {\rm log}_2|\mathcal {S}|}{{\rm log}_2|\mathcal {S}_i|}=\frac{1}{\alpha}=\frac{1}{d\beta}\leq \frac{k(2d-k+1)}{2dt}.$$
This completes the proof.
\end{IEEEproof}

This upper bound in the above theorem is achievable as the following Example \ref{Example_5row12column} to be shown. Particularly, we have to set $t\geq \frac{k(2d-k+1)}{2}$ to guarantee $\beta\geq 1$. Further, if $\beta=1$, the information rate $\rho_{\rm inf}$ is equal to $1/d$, which only depends on the size of repairing sets. Notice that this rate $1/d$ will degrade to 1 provided that the repairing function isn't considered, which is the optimal information rate for classical perfect secret sharing schemes (see \cite{Stinson-2006}). If it is allowed that the participant of failed share obtains some extra information during the repair process, i.e., $d\beta>\alpha$, then the repairing rate $\rho_{\rm rep}$ could decrease while the information rate $\rho_{\rm inf}$ would increase. Actually, there is a tradeoff between the repairing rate $\rho_{\rm rep}$ and the information rate $\rho_{\rm inf}$. At last, we indicate that if we apply the minimum bandwidth regenerating (MBR) codes (see \cite{Dimakis-etc-2010,Lin-Chung-2014,Rashmi-Shah-Kumar-2011}) in our construction of repairable-TSSSs, the constructed repairable-TSSSs can achieve the optimal repairing rate and the corresponding optimal information rate. In the following, we will take an example to design such an optimal $(n=4,k=2,d=3)$ repairable-TSSS with parameters $(\alpha=3, \beta=1)$ by our Construction.

\begin{eg}\label{Example_5row12column}
Consider a primitive polynomial $h(x)=x^5+x^2+1$ over the prime field $\mathbb{F}_{2}$. Let $\omega$ be a root of this primitive polynomial and further be a primitive element of the extended field $\mathbb{F}_{2^5}$. Furthermore, select a linearized polynomial:
$$f(x)=s x+\sum_{i=1}^{4} a_ix^{2^i}=\omega x+\sum_{i=1}^{4} x^{2^i}\in \mathbb{F}_{2^5}[x],$$
where, without loss of generality, let the secret be $s=\omega$. Then choose $5$ distinct elements $1$, $\omega$, $\omega^2$, $\omega^3$, $\omega^4$ in $\mathbb{F}_{2^5}$, evidently that are linearly independent over $\mathbb{F}_2$. Subsequently, compute $5$ evaluations:
\begin{align*}
&f(1)=\omega,\ f(\omega)=\omega^{19},\ f(\omega^2)=\omega^{20},\\
&\qquad f(\omega^3)=\omega^{25},\ f(\omega^4)=\omega^{22}.
\end{align*}
On the other hand, we use a $(4,2,3)$ MBR code with $(\alpha=3, \beta=1)$, whose public generator matrix is given by:
\begin{align*}
G=&[G_1|G_2|G_3|G_4]\\
=&\left[\begin{array}{ccc|ccc|ccc|ccc}
1 & 0 & 0  & 1 & 0 & 0  & 0 & 0 & 1  & 0 & 0 & 1 \\
0 & 1 & 0  & 0 & 0 & 0  & 1 & 0 & 1  & 0 & 0 & 1 \\
0 & 0 & 1  & 0 & 0 & 0  & 0 & 0 & 1  & 1 & 0 & 1 \\
0 & 0 & 0  & 0 & 1 & 0  & 0 & 1 & 1  & 0 & 0 & 1 \\
0 & 0 & 0  & 0 & 0 & 1  & 0 & 0 & 1  & 0 & 1 & 1
\end{array}\right].
\end{align*}
Each participant $P_i$ obtains his share
$$s_i=[f(1),f(\omega),f(\omega^2),f(\omega^3),f(\omega^4)]\cdot G_i,\ 1\leq i\leq 4,$$
that is,
\begin{align*}
&s_1=[\omega,\omega^{19},\omega^{20}],\ \ s_2=[\omega,\omega^{25},\omega^{22}],\\
&s_3=[\omega^{19},\omega^{25},\omega^2],\ s_4=[\omega^{20},\omega^{22},\omega^2].
\end{align*}
It is easy to verify that any 2 participants can recover the secret. As an example, we take the participants $P_2$ and $P_3$ into account. For $P_2$ and $P_3$, the following system of linear equations is true:
\begin{align}\label{eq:a6}
[s_2, s_3]=[f(1),f(\omega),f(\omega^2),f(\omega^3),f(\omega^4)]\cdot [G_2|G_3],
\end{align}
and specifically,
\begin{align*}
[\omega,\omega^{25},\omega^{22},\omega^{19},\omega^{25},\omega^2]=[\omega,\omega^{19},\omega^{20},\omega^{25},\omega^{22}] \cdot [G_2|G_3].
\end{align*}
Notice that the matrix $[G_2|G_3]$ is full row rank.

In addition, it follows:
\begin{align}
[s,a_1,a_2,a_3,a_4]\cdot X=[f(1),f(\omega),f(\omega^2),f(\omega^3),f(\omega^4)], \label{eq:a7}
\end{align}
where
\begin{align*}
X=&\begin{bmatrix}
1 & \omega       & \omega^2         & \omega^3         & \omega^4         \\
1 & \omega^2     & (\omega^2)^{2}   & (\omega^3)^{2}   & (\omega^4)^{2}   \\
1 & \omega^{2^2} & (\omega^2)^{2^2} & (\omega^3)^{2^2} & (\omega^4)^{2^2} \\
1 & \omega^{2^3} & (\omega^2)^{2^3} & (\omega^3)^{2^3} & (\omega^4)^{2^3} \\
1 & \omega^{2^4} & (\omega^2)^{2^4} & (\omega^3)^{2^4} & (\omega^4)^{2^4}
\end{bmatrix}\\
=&
\begin{bmatrix}
1 & \omega & \omega^2 & \omega^3 & \omega^4 \\
1 & \omega^2 & \omega^4 & \omega^6 & \omega^8 \\
1 & \omega^4 & \omega^8 & \omega^{12} & \omega^{16} \\
1 & \omega^8 & \omega^{16} & \omega^{24} & \omega \\
1 & \omega^{16} & \omega & \omega^{17} & \omega^2
\end{bmatrix}
\end{align*}
is a $5\times 5$ invertible matrix. Combining (\ref{eq:a6}) and (\ref{eq:a7}), $P_2$ and $P_3$ can obtain the following system of linear equations:
\begin{align}
&[s,a_1,a_2,a_3,a_4]\cdot X [G_2|G_3] \notag\\
=&[f(1),f(\omega),f(\omega^2),f(\omega^3),f(\omega^4)]\cdot [G_2|G_3] \notag\\
=&[\omega,\omega^{25},\omega^{22},\omega^{19},\omega^{25},\omega^{2}]. \label{eq:a8}
\end{align}
Since, as mentioned above, $X$ is invertible and $[G_2|G_3]$ is full row rank, $X \cdot [G_2|G_3]$ is still full row rank. Therefore, $P_2$ and $P_3$ can solve the secret $s=\omega$ from the system (\ref{eq:a8}) of linear equations (actually, solve all $s=\omega$, and $a_i=1$, $1\leq i \leq 4$). In other words, the recovery property is satisfied. Furthermore, by a simple observation, one can see that any failed share can be repaired by obtaining one encoded symbol from each of arbitrary other $3$ intact shares, i.e., $\beta=1$. For instance, suppose that the share $s_4$ fails, $P_1$, $P_2$, and $P_3$ provide $\omega^{20}$, $\omega^{22}$, and $\omega^{2}$ to $P_4$, respectively. Then $P_4$ recovers his share $s_4=[\omega^{20}, \omega^{22}, \omega^{2}]$. In addition, for every $1\leq i \leq 4$, it is not difficult to see that, for any $a\in \mathbb{F}_{2^5}$, the number of solutions, with the form $[a,a_1,a_2,a_3,a_4]$, of the linear equation $$[a,a_1,a_2,a_3,a_4]\cdot X G_i=s_i$$ is equal. This implies that the perfectly secure property is qualified for this repairable-TSSS.

Finally, for this repairable-TSSS, the repairing rate and the information rate of the scheme are respective:
\begin{align*}
\rho_{\rm rep}&=\frac{\alpha}{d\beta}=\frac{3}{3\times 1}=1,
\end{align*}
and
\begin{align*}
\rho_{\rm inf}&=\frac{k(2d-k+1)}{2dt}=\frac{1}{3},
\end{align*}
herein, $k=2,d=3,t=5$, which achieves the optimal repairing rate and the corresponding optimal information rate.
\end{eg}

\section{Conclusion}\label{Conclusion}
In this paper, we consider the repair problem of shares in threshold secret sharing schemes without the help of the dealer. We present a construction of repairable-TSSSs, which can accomplish the defined repairing function when a share fails and still guarantee all required properties of TSSSs, particularly, perfectly secure property. Finally, we analyze the performance of repairable threshold schemes by discussing the introduced repairing rate and traditional information rate, and indicate our proposed construction can obtain such optimal repairable-TSSSs by applying appropriate MBR codes.



\end{document}